\documentstyle[12pt]{article}

\newskip\humongous \humongous=0pt plus 1000pt minus 1000pt
\def\caja{\mathsurround=0pt}
\def\eqalign#1{\,\vcenter{\openup1\jot \caja
	\ialign{\strut \hfil$\displaystyle{##}$&$
	\displaystyle{{}##}$\hfil\crcr#1\crcr}}\,}
\newif\ifdtup

\def\oldreffmt#1{\rlap{[#1]} \hbox to 2\parindent{}}

\def\figfmt#1{\rlap{Figure {#1}} \hbox to 1in{}}

\def\beq{\begin{equation}}
\def\eeq{\end{equation}}

\relax

 1
\font\elevenrm=cmr10 scaled\magstep 1
 1

\renewenvironment{thebibliography}[1]
 { \elevenrm
   \begin{list}{\arabic{enumi}.}
    {\usecounter{enumi} \setlength{\parsep}{0pt}
     \setlength{\itemsep}{3pt} \settowidth{\labelwidth}{#1.}
     \sloppy
    }}{\end{list}}
\newlength{\abstractwidth}
\begin{document}
\thispagestyle{empty}
\pagestyle{empty}
\renewcommand{\thefootnote}{\fnsymbol{footnote}}
\renewcommand{\title}[1]{\vglue 1cm
\begin{center}\normalsize\bf #1\end{center}\par}
\renewcommand{\author}[1]{\vspace{2ex}{\normalsize\begin{center}
\setlength{\baselineskip}{3ex}#1\par\end{center}}}
\renewcommand{\thanks}[1]{\footnote{#1}} 
\renewcommand{\abstract}[1]{\vspace{4ex}\normalsize\begin{center}
\centerline{\bf Abstract}\par\vspace{2ex}\parbox{\abstractwidth}{{\small #1
\setlength{\baselineskip}{2.5ex}\par}}
\end{center}}
\newcommand{\starttext}{\newpage
\normalsize
\pagestyle{plain}
\setlength{\baselineskip}{14pt}\par
\setcounter{footnote}{0}
\renewcommand{\thefootnote}{\arabic{footnote}}}
\newcommand{\eqr}[1]{(\ref{#1})}
\newcounter{mysection}
\newcommand{\mysection}[1]{\stepcounter{mysection}
\setcounter{equation}{0}
\par\bigskip\noindent{\bf #1}\nopagebreak[4]\par\vskip .3cm}
\newcommand{\mysectionstar}[1]{                             
\par\medskip\noindent{\it #1}\nopagebreak[4]\par\vskip .3cm}%
\def\theequation{\themysection.\arabic{equation}}
%
\def\nl{&\cr\tablerule}
\newbox\bigbox
\setbox\bigbox=\hbox{\vrule height15pt depth8pt width0pt}
\def\bigstrut{\relax\ifmmode\copy\bigbox\else\unhcopy\bigbox\fi}
\def\sss{\scriptstyle}
\def\co{{\rm cosh}\,}
\def\si{{\rm sinh}\,}
\def\ta{{\rm tanh}\,}
\def\se{{\rm sech}\,}
\def\cs{{\rm csch}\,}
\def\e{{\rm exp}\,}
\def\lo{{\rm ln}\,}
\def\del{\partial}
\def\half{{\textstyle{1\over2}}}
\def\be{\begin{equation}}
\def\ee{\end{equation}}
\def\ch{\chi ^{a}}
\def\chh{\chi ^{a}_{H}}
\def\cs{\chi ^{a}_{S}}
\def\ic{\imath _{\chi}}
\def\dc{d_{\chi}}
\def\ds{d_{S}}
\def\lc{{\cal L}_{\chi}}
\def\ls{{\cal L}_{S}}
\def\lh{{\cal L}_{H}}
\def\ld{{\cal L}_{\dot\phi}}
\def\om{\omega _{ab}}
\def\pd{\dot\phi ^{a}}
\def\gab{g_{ab}}
\def\oab{\Omega _{ab}}
\def\rab{R_{ab}}
\def\gtt{g_{\theta\theta}}
\def\grt{g_{r\theta}}
\def\grr{g_{rr}}
\def\dt{\del _{\theta}}
\def\dr{\del _{r}}
\def\ax{{\vert x\vert}}
\def\ar{{\vert r\vert}}
\def\mint{\int ^{\infty}_{-\infty}}
\begin{flushright}
\begin{tabular}{r}
FERMI-PUB-92/383-T  \\
UMHEP--384
\end{tabular}
\end{flushright}
\vglue .2cm
\title{EXACT PATH INTEGRALS BY EQUIVARIANT LOCALIZATION}
\author{{Hans M. Dykstra\\}
\vglue .2cm
{\small\it Dept. of Physics and Astronomy\\
University of Massachusetts\\
Amherst, MA 01003}
}
\author{{Joseph D. Lykken and Eric J. Raiten\\}
\vglue .2cm
{\small\it Theory Group, MS106\\
Fermi National Accelerator Laboratory\\
P.O. Box 500, Batavia, IL 60510}
}

\date{}
\abstract{\small
It is a common belief among field theorists that path integrals can be
computed exactly only in a limited number of special cases, and that most
of these cases are already known. However recent developments, which
generalize the WKBJ method using equivariant cohomology, appear to contradict
this folk wisdom. At the formal level, equivariant localization would
seem to allow exact computation of phase space path integrals for an
arbitrary partition function! To see how, and if, these methods
really work in practice, we have applied them in explicit quantum mechanics
examples. We show that the path integral for the 1-d hydrogen atom,
which is not WKBJ exact, is localizable and computable using the more
general formalism. We find however considerable ambiguities in this
approach, which we can only partially resolve. In addition, we find a
large class of quantum mechanics examples where the localization procedure
breaks down completely.
}
\starttext
\mysection{\themysection  .\ Introduction}
The major failing of the path integral method applied to quantum
field theory or quantum mechanics is that most physically
relevant path integrals cannot be computed. The known exceptions fall
roughly into three categories. There are the Gaussian forms, which are
the basis of perturbation theory in quantum field theory. There are
also certain exactly solvable models, usually quantum mechanical
integrable systems. In such cases the path integral is often computable
due to the fact that the WKBJ approximation is exact. Thus the path
integral localizes to a sum over classical trajectories. The third
category is path integral representations of topological quantities
such as the Euler character or the Witten index. Here one often finds that
supersymmetry in the path integral implies an equivariant cohomology
structure, which localizes the path integral onto zero modes or
characteristic classes.

As emphasized in \cite{blau,moro,hie}, the common denominator
of most exactly computable path integrals is equivariant
cohomology and localization. This raises the possibility that by
studying these methods directly we may greatly expand our inventory
of doable path integrals, as well as WKBJ-like approximation schemes.
Though equivariant cohomology is
usually associated with supersymmetric theories, it actually
has very general application. The key ingredient required is
some sort of differential form structure among the physical, auxiliary,
or ghost variables. This is a rather weak requirement. In fact,
as recently shown by Blau, Keski-Vakkuri, and Niemi\cite{blau},
and by Niemi and Tirkkonen\cite{niemi}, equivariant cohomology
can be exhibited in an {\it arbitrary} phase space path integral.
Further, they showed formally that under seemingly weak conditions,
this results in equivariant localization
of these path integrals. They would thus be exactly computable!

Of course, we do not really expect arbitrary phase space
path integrals to be exactly computable.
Since equivariant localization proofs are very formal,
there are ample opportunities for subtle difficulties to creep in.
Also phase space path integrals are themselves rather disreputable.
As pointed out in \cite{blau}, there is a well known {\it false}
method for computing arbitrary phase space path integrals via
Hamilton-Jacobi theory and (invalid) canonical transformations
of the path integral variables\cite{schulman,dash}.

In short, there is no substitute for simple, explicit examples,
and that is the subject of this paper. In section 2.~we briefly
review equivariant cohomology and localization formulae for path
integrals. In section 3.~we make the formalism explicit by considering
the phase space path integrals for
the free particle and the harmonic oscillator. We use equivariant
localization to write the partition function of
a one-dimensional quantum mechanical
system with an {\it arbitrary} static potential
as an elementary contour integral. Along the way we
see that this derivation breaks down for essentially all potentials
which are bounded below.

In section 4.~we consider the 1-D hydrogen
atom, whose potential is not bounded below. Its path integral is
not WKBJ exact, and is not amenable to Morse theory analysis since
its classical trajectories all coalesce at two points in phase space.
Nevertheless we show that it is exactly computable via equivariant
localization, modulo an important ambiguity which we discuss.

\mysection{\themysection .\ Equivariant Cohomology and Localization}
Equivariant cohomology is a simple and powerful generalization of
ordinary de Rham cohomology and the calculus of differential forms
(for details, see \cite{bismut,berline}). To describe differential forms with
respect to some $d$-dimensional manifold $M$, it is convenient to
introduce a contravariant vector of anticommuting Grassmann variables
$c^a$. Then differential forms can be represented as covariant tensor
functions on $M$ contracted with the $c^a$'s:
\be
\eqalign{
0-{\rm forms}:\qquad &\phi (x) ,
\cr
1-{\rm forms}:\qquad &\phi _a(x)c^a ,
\cr
2-{\rm forms}:\qquad &\phi _{ab}(x)c^ac^b ,
\cr
{\rm etc. .}&\quad
\cr}
\ee
The exterior derivative operator $d$ can be written
\be
d=c^a{\del\over \del x^a} .
\ee
It takes $n$-forms to $n$$+$$1$-forms and is nilpotent: $d^2=0$.
De Rahm cohomology is the classification of closed forms (forms
which are annihilated by $d$) modulo exact forms (which are $d$
of something).

To define equivariant cohomology, we first introduce a new operation,
$\ic$, which is interior multiplication with respect to an
(arbitrary) vector $\ch (x)$. Explicitly:
\be
\ic = \int dc_a \ch
\ee
Clearly $\ic$ takes an $n$-form to an $n$$-$$1$ form:
\be
\eqalign{
&\ic \phi (x) = 0 ,
\cr
&\ic \phi _a(x)c^a = \phi _a \ch ,
\cr
&\ic \phi _{ab}(x)c^ac^b = \phi _{ab}( \ch c^b - \chi ^b c^a ) .
\cr}
\ee
Furthermore $\ic$ is nilpotent.

We can now define an equivariant exterior derivative $\dc$ by
\be
\dc = d + \ic .
\ee
Note that $\dc$ takes an $n$-form to both an $n$$+$$1$-form and
an $n$$-$$1$ form. Now, $\dc$ is not generally nilpotent:
\be
\dc ^2 = ( d\ic + \ic d ) = \lc ,
\ee
where $\lc$ is the Lie derivative with respect to the vector $\ch (x)$.
Thus $\dc$ is nilpotent only on the subspace of differential forms
which are annihilated by the Lie derivative $\lc$.

Without going into more rigorous mathematical detail, we are already
in a position to sketch the general derivation of equivariant localization
formulae. Suppose we want to compute an integral over the manifold
$M$ of some general form $\alpha$, where by a general form we mean
a linear combination of different $n$-forms (including a top-form):
\be
\int _M \alpha
\ee
In order to apply equivariant localization, we must be able to
find a vector $\ch$ such that, with respect to $\ch$, $\alpha$
is equivariantly closed:
\be
\dc \alpha = 0\;  .
\ee
In addition, we must be able to find some other differential form
$\beta$ which is in the subspace annihilated by
$\lc$: $\lc\beta = 0$, and is not equivariantly exact.
When this can be done, one constructs the following modified
integral:
\be
\int _M \alpha\; {\rm e}^{-\lambda \dc\beta} \; ,
\ee
where $\lambda$ is a numerical parameter. It is then straightforward
to show that, formally, this modified integral is independent
of the value of $\lambda$:
\be
\eqalign{
{d\over d\lambda}\int _M \alpha\; {\rm e}^{-\lambda \dc\beta}
&= -\lambda \int _M (\dc \beta )\alpha\; {\rm e}^{-\lambda\dc\beta}
\cr
&= \lambda \int _M \beta\; \dc\left( \alpha\; {\rm e}^{-\lambda
\dc\beta} \right)
\cr
&= 0 \; .
\cr}
\ee
In the above we used $\dc\alpha$$=$$0$, $\dc\dc\beta$$=$$0$,
and the fact that the integral of an equivariantly exact form
vanishes.

Thus, provided the limits $\lambda$$\rightarrow$$0$ and
$\lambda$$\rightarrow$$\infty$ exist, we obtain the
following localization formula:
\be
\int _M \alpha = \lim_{\lambda\to\infty} \int _M \alpha\;
{\rm e}^{-\lambda \dc\beta}\;  .
\label{lol}
\ee
The original integral over $M$ has been localized to a subspace
$M_\chi$, which is the support for the nontrivial equivariant
cohomology ($\dc\beta = 0$).

When applied to ordinary (finite dimensional) integrals over compact
manifolds, the above arguments can be made rigorous, and the right
hand side of \eqr{lol}
(for suitable choice of $\beta$) reduces to the WKBJ formula. This is the
Duistermaat-Heckman theorem.

At the formal level, this result was extended in \cite{blau}
to apply to general phase space path integrals. Consider a general
bosonic quantum mechanical phase space path integral:
\be
Z(T)=\int [d({\rm Liouville})]\; {\rm e}^{iS} =
\int [d\phi ^a]\sqrt{{\rm det}\Vert\om\Vert}
\e  i\int _0^T dt\left( \theta _a \pd - H(\phi ) \right) \; ,
\label{pspath}
\ee
where $\phi ^a(t)$ are the coordinates of phase space
(satisfying periodic boundary conditions),
$\theta _a(\phi )$ are their conjugates,
$H$ is the hamiltonian, and $\om$ is
an antisymmetric tensor giving the symplectic 2-form
whose matrix inverse defines Poisson brackets:
\be
\om = \del _a\theta _b - \del _b\theta _a \; .
\ee

Introducing real Grassmann auxiliary coordinates $c^a(t)$,
which also satisfy periodic boundary conditions, we
can write:
\be
Z(T) = \int [d\phi ^a][dc^a]\; \e i\int _0^T
dt \left( \theta _a \pd -H(\phi )
+\half c^a\om c^b \right) \; .
\ee
We can then identify
loop space differential forms as tensor functionals of
the $\phi ^a(t)$'s contracted with the $c^a(t)$'s.
There are two vectors of obvious significance for defining equivariant
operations. One is the hamiltonian vector field, which is defined
by the relation:
\be
\del _aH(\phi ) = \om \chi _H^b \; .
\label{hvf}
\ee
The other is $\pd$, since $\pd\del _a$ generates time translations on
our differential forms.

In \cite{blau}, equivariant cohomology is defined with respect
to the vector
\be
\cs (t) = \pd (t) - \chh (\phi (t))\;  ,
\label{chis}
\ee
since, as one can easily verify, $\ds S=0$.
Furthermore, the zeroes of $\cs (t)$ correspond to
the classical trajectories. Thus, if we shift the
action by $\ds$ of a 1-form $\beta$
\be
S\to S+\lambda \ds \beta ,
\ee
and choose $\beta$ proportional to $\cs$, we can expect to
obtain a WKBJ localization of $Z(T)$. Since both $\cs$ and $c^a$
are contravariant, we cannot construct such a $\beta$ without also
introducing a metric on loop space, $G_{ab}(t _1,t _2)$. Then we may
define $\beta$ as
\be
\beta = \int dt_1dt_2\; G_{ab}(t_1,t_2)\cs (t_1)c^b(t_2) \; .
\label{firstbeta}
\ee
Recalling that the localization argument requires $\ls \beta$$=$$0$,
one finds that this reduces to the condition
\be
\ls G_{ab}(t_1,t_2) = 0 \; .
\label{metcon}
\ee
In other words, $\cs$ must be a Killing vector of the loop space metric.
It is easy to see that $\ld$ automatically annihilates $G_{ab}(t_1,t_2)$
provided it takes the form
\be
G_{ab}(t_1,t_2) = \gab (\phi (t))\, \delta (t_1-t_2)\;  .
\label{goodmetric}
\ee
Then \eqr{metcon} reduces to the constraint
\be
\lh \gab (\phi ) = 0 \; .
\label{metconb}
\ee
In other words, $\chh$ must be a Killing vector of the finite dimensional
metric $\gab$.

As shown in \cite{blau}, the formal procedure just outlined results
in a WKBJ localization of the path integral \eqr{pspath}. Thus in
the case where these formal steps actually carry through, the path
integral is WKBJ exact. As already mentioned, there are a number of
known examples where this occurs. This is well-trodden territory,
and we have not attempted to explore it further.

Instead, we turn to the recent paper by Niemi and Tirkkonen\cite{niemi},
in which they apply a {\it different} equivariant localization scheme
to the general phase space integral \eqr{pspath}. The difference lies
in the choice of the 1-form $\beta$: instead of \eqr{firstbeta} they
choose
\be
\beta = \half \gab \pd c^b \; ,
\label{secondbeta}
\ee
where we suppressed the $t$ integrations which accompany
index contractions. The equivariant localization procedure is
otherwise the same as described above. One then obtains a localization
of $Z(T)$ not onto classical trajectories, but rather onto the time
independent modes of $\phi ^a(t)$ and $c^a(t)$. Their result
(correcting a minor typo) is
\be
Z(T) = \int d\phi _0^adc_0^a\; {\rm e}^{-iTH(\phi _0)
+i\half Tc_0^a\om c_0^b}
\;\sqrt{{\rm Det}\left[ {\half (\oab + \rab )\over
\si \left[ {T\over 2}(\oab + \rab )\right]} \right] }.
\label{ntformula}
\ee
In this expression, the determinant is an ordinary matrix
determinant, and the integrals are finite dimensional integrals
over the time-independent real variables $\phi _0^a$ and real
Grassmann variables $c_0^a$. The tensors $\oab$ and $\rab$
can be expressed as geometric quantities in terms of the metric
$\gab$ and the hamiltonian vector $\chh$ which is a Killing
vector of $\gab$:
\be
\eqalign{
\oab &= 2\chi ^H_{a;b} \quad ,
\cr
\rab &= R_{abcd}c_0^cc_0^d \quad .
\cr}
\ee

Now \eqr{ntformula} is a remarkably simple result. If this
localization proved to be valid for some large class of physically
relevant partition functions, it would be an exciting and important
achievement. Less optimistically, we have here at the very least
an alternative to WKBJ methods which is equally nonperturbative.
Further, both WKBJ and the Niemi-Tirkkonen formula are merely examples
of much more general equivariant localization techniques.

\mysection{\themysection .\ Application to
1-D Quantum Mechanics with Static Potentials}
The simplest possible application of \eqr{ntformula} is to
one dimensional quantum mechanics with static potentials. In this
case there are only two phase space coordinates, $x(t)$ and $p(t)$,
and the Hamiltonian is of the form
\be
H = \half p^2 + V(x) .
\ee
For the moment,
let us confine ourselves to potentials which are bounded below.
Then it is convenient to define ``harmonic'' and polar coordinates
as follows:
\be
\eqalign{
p&= r{\rm sin}\theta ,
\cr
V(x) - V_{\rm min} &= \half\omega ^2 y^2 =  r{\rm cos}\theta ,
\cr
H &= \half p^2 + \half \omega ^2 y^2
+V_{\rm min} = \half r^2 + V_{\rm min} .
\cr}
\ee
Note that we only change variables in the time independent coordinates
of the localized integral, never in the original phase space path integral.

The advantage of working in polar coordinates is that the
hamiltonian vector field has only one nonvanishing component:
\be
\chi ^\theta = \chi = -\omega {dy\over dx} \; ,
\ee
and it is straightforward to solve the constraint $\lh \gab = 0$,
which is equivalent to:
\be
\eqalign{
\chi \dt\gtt + 2\gtt\dt\chi &= 0 \; ,
\cr
\dt (\chi\grt ) + \gtt\dr\chi &= 0 \; ,
\cr
\chi\dt\grr + 2\grt\dr\chi &= 0 \; .
\cr}
\ee
The general solution is:
\be
\eqalign{
\gtt &= {f(r)\over \chi ^2} \; ,
\cr
\grt &= {f(r)\over \chi ^2}\int d\theta \;\dr\left({1\over \chi}\right)
+ {f_2(r)\over \chi} \; ,
\cr
\grr &= {\chi ^2 \over f(r)}(\grt )^2 + f_3(r) \; .
\cr}
\ee
Note that the metric is not completely determined by the constraint;
the general solution (3.5) involves three arbitrary functions
of $r$: $f$, $f_2$, and $f_3$. This is not surprising, since we expect
that the localized integral \eqr{ntformula} is at least partially
metric independent.

A much worse feature of the general solution (3.5)
is that $\grt (r,\theta )$ is not, in general, a single-valued
function. To see this, observe that
$\grt(\theta$$=$$0) = \grt(\theta$$=$$2\pi )$ only obtains if
\be
\dr \int _0^{2\pi} {1\over \chi} = 0 \;,
\ee
which is equivalent to
\be
\int _0^{2\pi} d\theta \; {dx\over dy} = {\rm constant} \; .
\label{sing}
\ee
The harmonic oscillator gives the only solution
of \eqr{sing}. Thus we conclude that, for essentially all static
potentials which are bounded below, the Niemi-Tirkkonen
equivariant localization procedure fails, due to the nonexistence
of a single-valued metric satisfying the Lie derivative constraint.

What of the harmonic oscillator, the sole survivor of this calamity?
In this case we have
\be
\eqalign{
\chi = -\omega \; &,
\cr
\gtt = {f(r)\over \omega ^2} \; ,\;
\grt = -{f_2(r)\over \omega} \; &,\; \grr =
{f_2^2\over f}+f_3(r) \; .
\cr}
\ee
The antisymmetric tensor $\oab$ and the curvature tensor $R_{abcd}$
each have only one independent component:
\be
\Omega _{\theta r} = -{f^\prime (r)\over \omega}  \; ,\quad
R_{\theta r\theta r} = {-c^\prime (r)\over 2\omega c(r)}
\Omega _{\theta r} \; ,
\ee
where we have introduced the (as yet) arbitrary function $c(r)$:
\be
c(r) = {\Omega _{\theta r}
\over 2\sqrt{g}} = -{f^\prime \over 2\sqrt{ff_3}} \; .
\ee
It is now straightforward to plug into the localization formula
\eqr{ntformula}. Performing the Grassmann integrals and the $\theta$
integral leads to the following simple result:
\be
Z(T) = {1\over \omega}\int_0^\infty  dr\; {d\over dr}
\left( {c\over {\rm sin}\, cT}{\rm e}^{-i\half r^2 T} \right) \; .
\ee
So, in the special case of the harmonic oscillator, the
localization formula reduces to a total derivative. The final
result is
\be
Z(T) = {1\over \omega} {c(0)\over {\rm sin}[c(0)T]} \; .
\ee

It is surprising that, even in such a simple case as the harmonic
oscillator, the final result of the localization procedure is
ambiguous, i.e. it depends on the value at $r$$=$$0$ of the
arbitrary function $c(r)$. The correct result for the path integral
is only obtained if we impose the additional ad hoc boundary
condition
\be
\lim_{r\to 0}c(r) = \half \omega \; .
\ee
As we will see in the next section, such ambiguities appear to
be a general feature of equivariant localization of
phase space path integrals. Thus the localization formulae,
even when they are otherwise valid, are incomplete.
This could have been anticipated even in the formal derivation.
An obvious requirement\cite{pas}
which must be incorporated into the
localization procedure is that the 1-form $\beta$ should be
homotopically equivalent to $0$ under the ``supersymmetry''
transformation generated by $\ds$. Thus one expects generally that
additional inputs are required when choosing $\gab$, in order to
ensure that we are in the trivial homotopy class.

Before moving on to more interesting problems, let us consider
the phase space path integral for the free particle. In this case
$H=\half p^2$, and $\chi ^x$$=$$\chi$$=$$p$. The localization
procedure again has ambiguities, which can be eliminated by
fixing the metric for the free particle to be:
\be
\gab = \left(\matrix{g_{xx}&g_{xp}\cr
                     g_{px}&g_{pp}\cr}\right)
=\left(\matrix{0&0\cr 0&1\cr}\right) \; .
\label{fpmet}
\ee
The localization formula can be reduced to
\be
Z(T) = \int dpdx\; {1\over p} {d\over dp} \left(
{c(p)\over {\rm sin}\, cT} {\rm e}^{-i\half p^2T}  \right)\; .
\ee
With the metric \eqr{fpmet} we have $c(p)$ vanishing identically,
and the above produces the correct result.

\mysection{\themysection .\ The 1-D Hydrogen Atom}
The 1-d hydrogen atom is the static potential problem with
hamiltonian\cite{lou}
\be
H = \half p^2 - {e^2\over \ax} \; .
\ee
While the partition function can be computed exactly by
solving the Schrodinger equation, the path integral for the
partition function cannot be directly performed by standard
techniques. In particular it is not WKBJ exact. The classical
bound state orbits coalesce at $x$$=$$0$, $p\to \pm\infty$,
making this system highly unsuitable for localization onto
classical trajectories. We wish to examine whether another
form of equivariant localization, such as \eqr{ntformula},
may apply to systems of this type.

In considering bound states it will again be useful to change
variables (of the time independent modes only!). We define
hyperbolic coordinates
\be
\eqalign{
p &= \ar \si \alpha \quad ,
\cr
x &= {2e^2\over r\ar \co ^2\alpha} \quad ,
\cr}
\ee
so that $H= -\half r^2$.
The original phase space maps into $-\infty \leq r,\alpha \leq \infty$.
The hamiltonian vector has only one nonvanishing component:
\be
\chi ^\alpha = \chi = -{1\over 4e^2}r^3\co ^3\alpha \; .
\ee
The Lie derivative constraint $\lh\gab = 0$ has exactly the same
form as (3.4), with $(r,\theta )$ replaced by $(r,\alpha )$.
Thus the general solution for $\gab$ has precisely the
form (3.5). However
in the present case, since $\alpha$ is a hyperbolic coordinate,
we do not encounter a single-valuedness problem in defining
$g_{r\alpha}$. Explicitly:
\be
g_{r\alpha} = {12e^2f(r)\over r^4\chi}\left[ {\si\alpha\over 2\co ^2\alpha}
+\half {\rm tan}^{-1}(\si\alpha ) \right] + {f_2(r)\over \chi} \; ,
\ee
which is perfectly well-defined.

A trivial calculation gives
\be
\Omega _{\alpha r} = {f^\prime (r)\over \chi} \; .
\ee
The curvature can then be computed rather easily by exploiting the
relation
\be
\Omega _{\alpha r;r} = 2R_{\alpha r\alpha r}\chi \; .
\ee
One obtains
\be
R_{\alpha r\alpha r} = {1\over 2\chi ^2}\left( f^{\prime\prime}
-{(f^\prime )^2\over 2f}-{f^\prime f_3^\prime \over 2f_3 }\right) \; .
\ee
With these results, one can easily show that the localization
formula \eqr{ntformula} can be reduced to the following simple
expression:
\be
Z(T) = \mint dr \mint d\alpha \; {1\over \chi}{d\over dr}
\left( {c\over {\rm sin}\, cT} {\rm e}^{i\half r^2T} \right) \; ,
\ee
where the function $c(r)$ is defined by
\be
c(r) = {f^\prime \over 2\sqrt{ff_3}} \; .
\label{magic}
\ee
In fact \eqr{magic} is a general result; it follows from
\eqr{ntformula} for any static potential which is unbounded
below.

As for the harmonic oscillator, we have obtained a remarkably simple
result, which is unfortunately ambiguous due to its dependence
on the (as yet) arbitrary function $c(r)$. Actually the ambiguity
appears to be much worse, since the harmonic oscillator result
only depended on $c(0)$. We can improve the situation by performing
some additional manipulations on \eqr{magic}. We perform the
$\alpha$ integration, and expand sin$\, cT$ as a power series.
Then, assuming that we can interchange integrations and summations,
we have a simple expression for the integrated partition function:
\be
\int _0^\infty dT\; Z(T) = -12\pi e^2 \sum _{n=1}^{\infty}
\mint dr\; {1\over r^4} {1\over (2n-1) - \left( {r^2\over 2c(r)} \right)}
\; .
\label{magb}
\ee
We then make
the mild assumption that the only singularities of the integrand
in \eqr{magb}, assumed to be
an analytic function of complex $r$, are poles on the real axis.
We can then convert \eqr{magb} into a contour integral, adding
the prescription that the contour includes all poles except the
one at $r=0$ (if we include all the poles, the integral simply
vanishes). However, since the contour can be closed above or below,
this is the same as including {\it only} the pole at $r=0$.
We therefore obtain
\be
\int _0^\infty dT\; Z(T) = 4\pi ^2 e^2 i \sum _{n=1}^{\infty}\;
\lim_{r\to 0}{d^3\over dr^3}\left( {1\over (2n-1) -
\left( {r^2\over 2c(r)} \right) } \right)  \; .
\label{magc}
\ee

Thus we see that the result depends only on the behavior
of $c(r)$ as $r$ goes to zero.
To resolve this remaining ambiguity, we need an additional
input. To see what this is, let us consider the main difference
between the WKBJ localization of path integrals and the localization
procedure which led to \eqr{ntformula} and \eqr{magic}. In the
former case the original path integral is reduced to a sum of
small fluctuation integrals around the classical trajectories.
In the latter case, the original path integral is reduced to
a sum of small fluctuation integrals around individual points
in the phase space parametrized by the time independent
modes $(x,p)$. From this fact we conclude that the contribution
to \eqr{magic} from the region $x\to\infty$, $p\sim 0$, should
be approximately equal to the corresponding contribution in
(3.39) for the free particle, provided that $T$ is not near $\infty$.

We have applied this matching condition by computing the general
metric for the 1-d hydrogen atom in $(x,p)$ coordinates.
We then matched to the free particle metric \eqr{fpmet}
to leading order in large $x$, and to leading and next-to-leading
order in small $p^2x$. This is sufficient to give the leading order
behavior of $c(r)$ for small $r$:
\be
c(r) = {r^3\over 2e^2}\left( 1 + c_1 r + c_2 r^2 +\ldots \right) \; .
\ee

Actually, the evaluation of \eqr{magc} requires knowledge of
the corrections $c_1$ and $c_2$. While these can be computed
in principle via perturbation theory, we have not done so.
Regardless, the leading order behavior is enough to see that
\eqr{magc} reproduces the correct form for the bound state
spectrum of the 1-d hydrogen atom:
\be
E_n \propto {e^4\over n^2} \quad .
\ee

We consider this a nontrivial success of equivariant localization
techniques applied to physically relevant path integrals.
While we are still operating at a rather primitive level
compared to standard WKBJ, we believe that the potential
for obtaining new results via
these more general methods of equivariant localization
is much greater.

%
\def\cmp#1{{\it Comm. Math. Phys.} {\bf #1}}
\def\cqg#1{{\it Class. Quantum Grav.} {\bf #1}}
\def\plb#1{{\it Phys. Lett.} {\bf #1B}}
\def\pl#1{{\it Phys. Lett. } {\bf #1B}}
\def\prl#1{{\it Phys. Rev. Lett.} {\bf #1}}
\def\prd#1{{\it Phys. Rev.} {\bf D#1}}
\def\prr#1{{\it Phys. Rev.} {\bf #1}}
\def\prb#1{{\it Phys. Rev.} {\bf B#1}}
\def\npb#1{{\it Nucl. Phys.} {\bf B#1}}
\def\ncim#1{{\it Nuovo Cimento} {\bf #1}}
\def\jmp#1{{\it J. Math. Phys.} {\bf #1}}
\def\aam#1{{\it Adv. Appl. Math.} {\bf #1}}
\def\mpl#1{{\it Mod. Phys. Lett.} {\bf A#1}}



\begin{thebibliography}{9}
\bibitem{blau} M. Blau, E. Keski-Vakkuri, and A. Niemi, \plb{246} (1990) 92.
\bibitem{moro} A. Morozov, A. Niemi, and K. Palo, \plb{271} (1991) 365.
\bibitem{hie} A. Hietam\"aki, A. Morozov, A. Niemi, and K. Palo,
\plb{263} (1991) 417.
\bibitem{niemi} A. Niemi and O. Tirkkonen, \plb{293} (1992) 339.
\bibitem{schulman} L. Schulman, ``Techniques and Applications of Path
Integration'', John Wiley \& Sons (New York) 1981.
\bibitem{dash} R. Dashen, B. Hasslacher, and A. Neveu, \prd{12} (1975) 2443.
\bibitem{bismut} J-M Bismut, \cmp{103} (1986) 127.
\bibitem{berline} N. Berline, E. Getzler, and M. Vergne,
``Heat Kernels and Dirac Operators'', Springer-Verlag (Berlin) 1992.
\bibitem{pas} A. Niemi and P. Pasanen, \plb{253} (1991) 349.
\bibitem{lou} R. Loudon, {\it Am. J. Phys.} {\bf 27} (1959) 649.
\end{thebibliography}
\end{document}